\documentclass[12pt,twocolumn]{iopart}
\usepackage{graphicx}

\begin{document}

\title[Growth of Bi$_2$(Sr$_{2-\emph{x}}$La$_\emph{x}$)CuO$_{6+\delta}$ Single Crystals]{Growth, Characterization
and Physical Properties of High$-$Quality Large Single Crystals of
Bi$_2$(Sr$_{2-\emph{x}}$La$_\emph{x}$)CuO$_{6+\delta}$ High
Temperature Superconductors}

\author{Jianqiao Meng, Guodong Liu, Wentao Zhang, Lin Zhao, Haiyun Liu, Wei Lu, Xiaoli Dong and X. J. Zhou$^{*}$}

\address{National Laboratory for Superconductivity, Beijing National Laboratory for
Condensed Matter Physics, Institute of Physics, Chinese Academy of
Sciences, Beijing 100190, China} \ead{XJZhou@aphy.iphy.ac.cn}

\begin{abstract}\\
High quality large
Bi$_2$(Sr$_{2-\emph{x}}$La$_\emph{x}$)CuO$_{6+\delta}$ (La-Bi2201)
single crystals have been successfully grown by the traveling
solvent floating zone technique. The samples are characterized by
compositional and structural analyzes and their physical properties
are investigated by magnetic susceptibility and resistivity
measurements. Superconducting samples with sharp superconducting
transitions are obtained covering a wide range of doping from
overdoped (x$<$0.40), optimally-doped (x$\sim$0.40), underdoped
(0.40$<$x$<$0.84) to heavily underdoped without superconducting
transition (x$>$0.84). Crystals as large as $\sim$40 $\times$ 2.0
$\times$ 1 mm$^3$ are obtained for x=0.73. Sharp superconducting
transition with a width less than 2 K and nearly perfect Meissner
signal of superconductivity are achieved for x=0.40. The
availability of the La-Bi2201 single crystals will provide an ideal
system to study the physical properties,  electronic structure and
mechanism of high temperature superconductivity.

\end{abstract}

\maketitle

\section{Introduction}

The physical properties of the copper$-$oxide (cuprate)
superconductors are characterized by the unusually high critical
temperature (T$_c$), the unusual superconducting state with
predominantly {\it d}-wave pairing \cite{Tsuei}, and exotic normal
state where a pseudogap is present in the underdoped and optimally
doped regions\cite{Timusk}. Since the conventional Fermi liquid
theory for metals\cite{FermiLiquid} and the BCS theory for
conventional superconductivity\cite{BCSTheory} can not be directly
applicable to the cuprate superconductors, new theories have to be
developed. Nearly twenty years after its first
discovery\cite{Bednorz},  the mechanism of high-temperature
superconductivity in cuprates remains an outstanding issue in
condensed matter physics\cite{Orenstein,DABonn}. Critical
experiments such as angle-resolved photoemission
spectroscopy(ARPES)\cite{ARPES}, neutron scattering\cite{Neutron},
Scanning tunneling microscopy/spectroscopy(STM/STS)\cite{Fischer}
and etc. are necessary in stimulating, checking and developing new
theories. High quality single crystal samples are crucial for
carrying out these experiments. In some experiments,  large size and
large quantity of single crystals are needed such as in neutron
scattering measurements.

The Bi$_2$Sr$_2$CuO$_6$ (Bi2201) superconductor is an ideal system
in investigating high temperature superconductivity in
cuprates\cite{A.Maeda}. First, it has a layered structure which can
be easily cleaved to get a clean and smooth surface, which is
necessary for experiments like ARPES and STM/STS. Second, it has
simple single-layered crystallographic structure with one CuO$_2$
plane within two adjacent block layers, this gives single band and
single Fermi surface sheet, avoiding possible complications from
two-layered or multi-layered compounds such as bilayer splitting in
double-layered Bi$_2$Sr$_2$CaCu$_2$O$_8$
(Bi2212)\cite{BiLayerSplitting}. Third, by partially replacing
Sr$^{2+}$ with some rare earth ions (R$^{3+}$) in
Bi$_2$(Sr$_{2-x}$R$_x$)CuO$_{6+\delta}$, the critical temperature of
Bi2201 can be increased from around $\sim$9 K for the pristine
Bi2201\cite{B.Liang2,Huiqian.Luo1} to nearly 38 K for R=La
(lanthanum)\cite{S.Sastry,Eisaki,Y.Ando3}. The relatively lower
critical temperature, compared with that of Bi2212 and
YBa$_2$Cu$_3$O$_{7-\delta}$ with a maximum T$_c$ higher than 90 K,
is beneficial in investigating normal state properties while
suppressing possible thermal broadening caused by high temperature.
Fourth, by substituting Sr$^{2+}$ with rare-earth (R$^{3+}$) which
introduces electrons, together with oxygen annealing, the doping
level of the samples can be varied over a wide range. In the case of
La-doped Bi2201, various doping levels from underdoped, optimally
doped and overdoped samples, and even heavily underdoped
non-superconducting samples can be
obtained\cite{B.Liang2,Huiqian.Luo1,Y.Ando1,Schloegl,Khasanova}.
This large doping range is very important for addressing many
important issues such as metal$-$insulator transition,
non-superconducting$-$superconductor transition and other doping
evolution of the electronic structure and physical properties of
cuprate superconductors.

There have been a few attempts in growing
Bi$_2$(Sr$_{2-x}$La$_x$)CuO$_{6+\delta}$ (abbreviated as La-Bi2201
hereafter) single
crystals\cite{N.L.Wang,W.L.Yang,B.Liang1,S.Ono1,Huiqian.Luo2}. While
the usual self$-$flux method\cite{N.L.Wang,W.L.Yang} is possible,
the traveling solvent floating zone method is advantageous in
getting large size, better control of the composition and high
quality by avoiding possible contaminations from
crucibles\cite{B.Liang1,S.Ono1,Huiqian.Luo2}. In this paper, we
report growth of high quality large-size La-Bi2201 single crystals
by traveling solvent floating zone method. The single crystals we
have grown cover a wide doping range from overdoped (x=0.25), to
optimally doped (x=0.40), to underdoped (x larger than 0.40) all the
way to heavily underdoped without superconducting transition
(x=0.84). The crystal size we have obtained is up to 3$\sim$4cm
long. The superconducting samples show a sharp supercondcting
transition and nearly a perfect Meissner shielding of
superconductivity. The successful growth of these single crystal
will be very helpful in carrying out some critical experiments to
investigate high temperature cuprate superconductors.

\section{Experiment method}

The Bi$_2$(Sr$_{2-x}$La$_{x}$)CuO$_{6+\delta}$ single crystals are
grown by traveling solvent floating zone method using an infrared
radiation furnace (\emph{Crystal Systems Inc.}) equipped with four
300 W halogen lamps. The high temperature gradient required to
stabilize the molten zone can be achieved by ellipsoidal mirrors
that focus the image of the four lamps at a common position. The
temperature of the molten zone can be controlled precisely by
adjusting the output power. The crystal growth was carried out in an
enclosed quartz tube where the growth atmosphere can be controlled.

\begin{table}
\caption{Nominal and measured composition, growth
conditions,\emph{c}-axis lattice constant and \emph{T$_{c,onset}$}
of Bi$_2$(Sr$_{2-\emph{x}}$La$_\emph{x}$)CuO$_{6+\delta}$ single
crystals} \scriptsize\rm
\begin{tabular}{llcllc}
\br Nominal Composition & Measured Composition & Growth Rate(mm/h)&
Growth Atmosphere&\emph{c}({\AA})&\emph{T}$_{c,onset}$
(K)\\
\mr
Bi$_2$Sr$_{1.75}$La$_{0.25}$CuO$_{6+\delta}$&Bi$_2$Sr$_{1.68}$La$_{0.30}$CuO$_{6+\delta}$&0.5
&100 cc/min Air flow&24.519&28\\[0.5ex]
Bi$_2$Sr$_{1.63}$La$_{0.37}$CuO$_{6+\delta}$&Bi$_2$Sr$_{1.72}$La$_{0.36}$CuO$_{6+\delta}$&0.5
&100 cc/min O$_2$ flow&24.445&30\\[0.5ex]
Bi$_2$Sr$_{1.60}$La$_{0.40}$CuO$_{6+\delta}$&Bi$_2$Sr$_{1.64}$La$_{0.39}$CuO$_{6+\delta}$&0.5
&100 cc/min O$_2$ flow&24.392&32\\[0.5ex]
Bi$_2$Sr$_{1.40}$La$_{0.60}$CuO$_{6+\delta}$&Bi$_2$Sr$_{1.31}$La$_{0.64}$CuO$_{6+\delta}$&0.5
&50 cc/min O$_2$ flow&24.222&24.5\\[0.5ex]
Bi$_2$Sr$_{1.30}$La$_{0.70}$CuO$_{6+\delta}$&Bi$_2$Sr$_{1.28}$La$_{0.71}$CuO$_{6+\delta}$&0.5
&20 cc/min O$_2$ flow&24.171&20\\[0.5ex]
Bi$_2$Sr$_{1.27}$La$_{0.73}$CuO$_{6+\delta}$&Bi$_2$Sr$_{1.29}$La$_{0.75}$CuO$_{6+\delta}$&0.5
&20 cc/min O$_2$&24.148&18\\[0.5ex]
Bi$_2$Sr$_{1.16}$La$_{0.84}$CuO$_{6+\delta}$&Bi$_2$Sr$_{1.10}$La$_{0.97}$CuO$_{6+\delta}$&0.5
&20 cc/min O$_2$&24.052&ns \\\hline\br
\end{tabular}
\end{table}

Feed rods and seed rods are first prepared by conventional
solid-state reaction method before growing single crystals. Starting
materials of Bi$_2$O$_3$, SrCO$_3$, La$_2$O$_3$ (pre-heated for 5
hours at 860$^\circ$C to remove the adsorbed water and CO$_2$) and
CuO with 99.99$\%$ purity were weighed according to the chemical
formula Bi$_2$(Sr$_{2-x}$La$_x$)CuO$_{6+\delta}$ with different x
and mixed in an agate mortar for about 8 hours. The mixed powder was
calcined at 780$^\circ$C for 24 hours and the calcined product was
reground into fine powders. This calcination and grinding procedure
was repeated for four times to ensure a complete reaction and
homogeneity of the calcined powders. The calcined powder was pressed
into a cylindrical rod of 7 mm in diameter and 120 mm in length
under a hydrostatic pressure of $\sim$ 70 MPa, followed by sintering
in a vertical molisili furnace at 870$^\circ$C for 48 h in air.  The
sintered rod was then pre-melted in the floating zone furnace at a
traveling velocity of 25$\sim$30 mm/hour to obtain a dense feed rod
with $\sim$6 mm in diameter and $\sim$120 mm  in length.  The
pre-melting is a crucial step in achieving a stable molten zone
during the crystal growth because it may prevent the molten zone
from penetrating into the feed rod. An ingot about 1.5 cm long cut
from the pre-melted feed rod was used as a seed rod.

\begin{figure}[tbp]
\begin{center}
\includegraphics[width=0.95\columnwidth,angle=0]{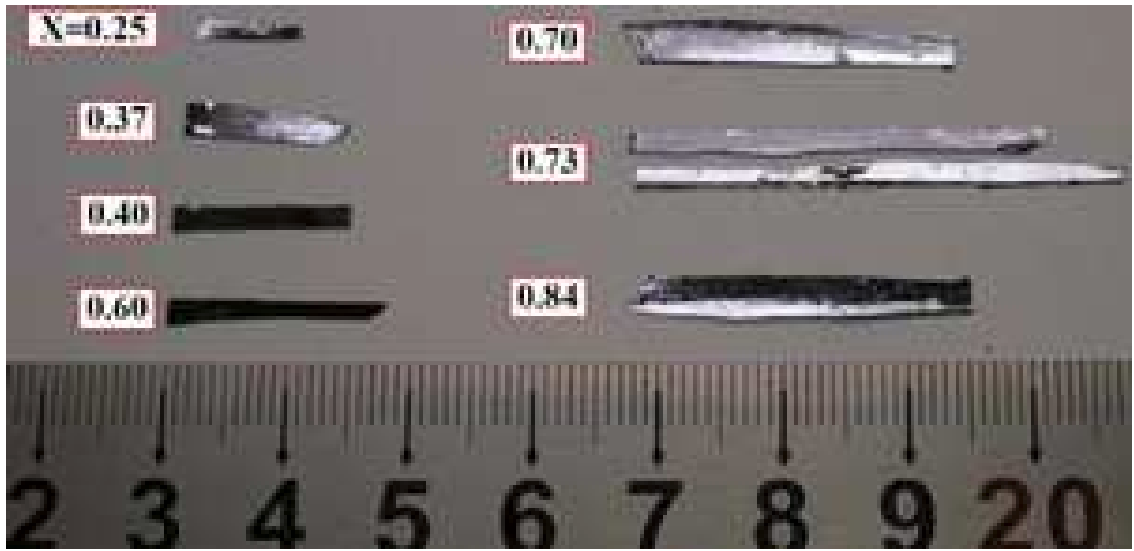}
\end{center}
\caption{Photos of Bi$_2$(Sr$_{2-x}$La$_x$)CuO$_{6+\delta}$ single
crystals cleaved from as-grown ingots with various nominal
compositions: \emph{x} =0.25,0.37, 0.40, 0.60, 0.70, 0.73 and 0.84.
Some crystals with large x can reach 3$\sim$4 cm in length.
 }
\end{figure}

\begin{figure}[tbp]
\begin{center}
\includegraphics[width=0.95\columnwidth,angle=0]{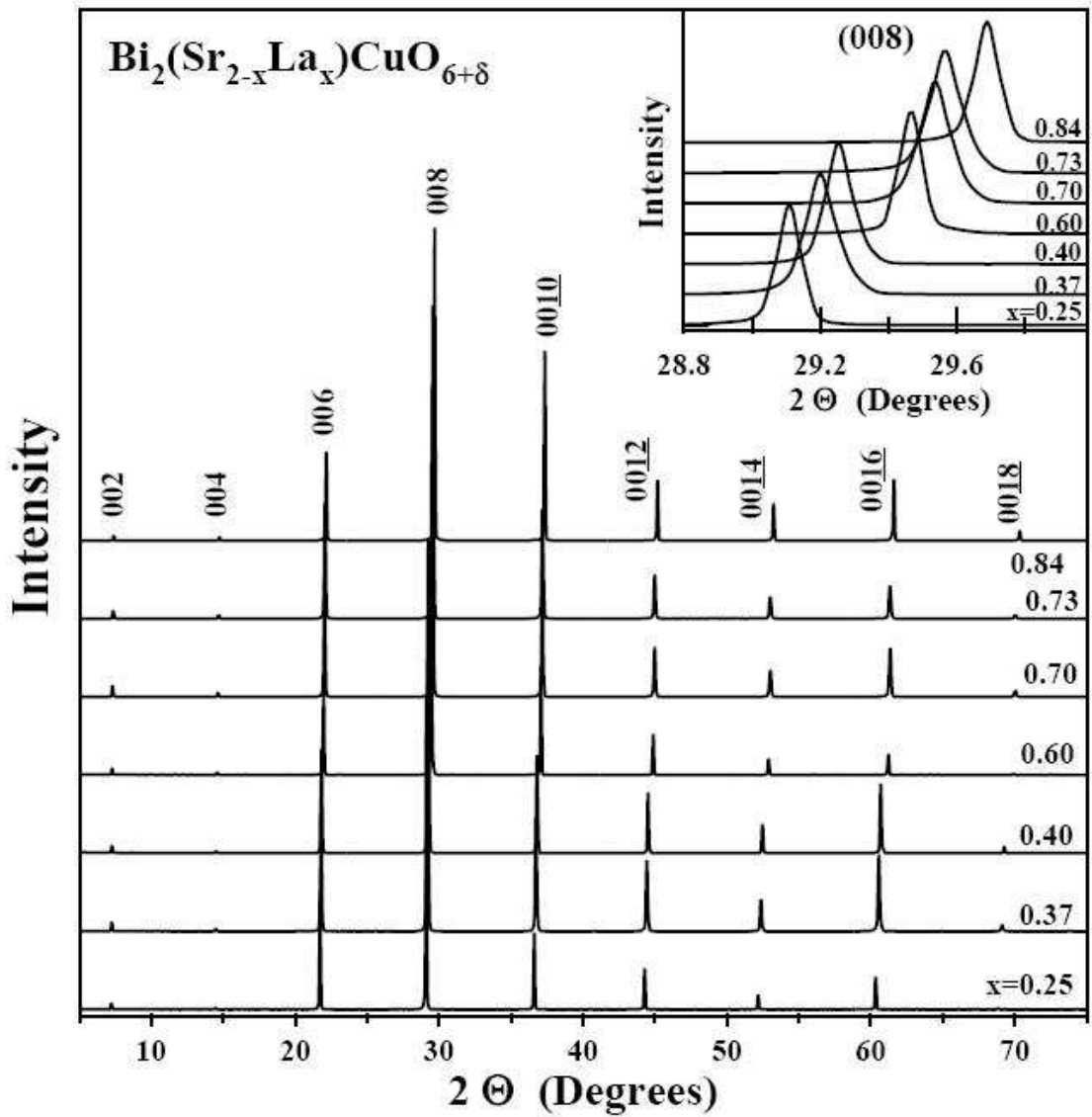}
\end{center}
\caption{XRD patterns for cleaved single crystal with various La
concentration (\emph{x}).
 }
\end{figure}

The crystal growth involves many trials to optimize growing
conditions, including the power used, growth rate, rotation rates of
feed rod and the seed rod, and growing atmosphere. These conditions
vary for different compositions. Table 1 lists La-Bi2201 crystals
with different compositions, their growth conditions, and the
structural and superconducting properties.  Some key parameters to
obtain high-quality large La-Bi2201 single crystals are briefly described below:\\
(1). The output power is in principle dictated by the melting point
of the feed rod. In practice it can be estimated during the
pre-melting process. An optimized output power can be achieved by
fine-tuning within a narrow range which gives rise to an appropriate
molten zone that can stay stable during the entire growth process.\\
(2). The growth rate is determined by the atomic diffusion kinetics
during the crystal growing process which may vary significantly
among different materials. The stability of the molten zone and the
crystal size are other factors to take into account.  For La-Bi2201,
when a lower growth rate ($<$0.45 mm/hour) was used, the molten zone
tends to become unstable causing frequent collapse of the zone. When
a higher growth rate ($>$0.8 mm/hour) is used, although the molten
zone gets more stable, the size of the single crystals size gets
smaller. As a compromise, a growth rates of 0.5 mm/hour was used for
growing La-Bi2201 single crystals.\\
(3). The feed rod and the seed rod both rotate in the opposite
direction in order to ensure efficient mixing and uniform
temperature distribution in the molten zone.  A 30/20 rpm rate was
used for feed rod (upper shaft)/seed rod (lower shaft) in growing
La-Bi2201.  The upper shaft rotates faster to make the molten zone
more homogenous, while the lower shaft with a slower rate is to keep
the molten zone stable. It is important to make sure that there is
no deflection between the seed rod and feed rod during their rotations.\\
(4). The growth atmosphere has an important effect on the stability
of the molten zone and the crystal quality. Depending on the
composition, we used both air and oxygen atmosphere (Table 1). We
note that when oxygen pressure is used, the pressure should be no
higher than 2 Bar. Otherwise there would be bubbles appearing in the
molten zone that cause it unstable. The gas flowing rate is an
important factor for affecting the crystal size; large La-Bi2201 crystals
are obtained when 20$\sim$100 cc/min oxygen or air flow rates is used,
as listed in Table 1 for different compositions. \\

Fig.1 shows some typical La-Bi2201 single crystals with different La
concentration that were cleaved from the top of the as-grown ingots.
We note that for the initial part of the ingot (0$\sim$3 cm),
because the growing conditions were not yet optimized and stable,
the crystal size is usually small with lower quality. Large
high-quality crystals are obtained in the final part of the ingot
when the growth process is optimized and stable.

\begin{figure}[tbp]
\begin{center}
\includegraphics[width=0.8\columnwidth,angle=0]{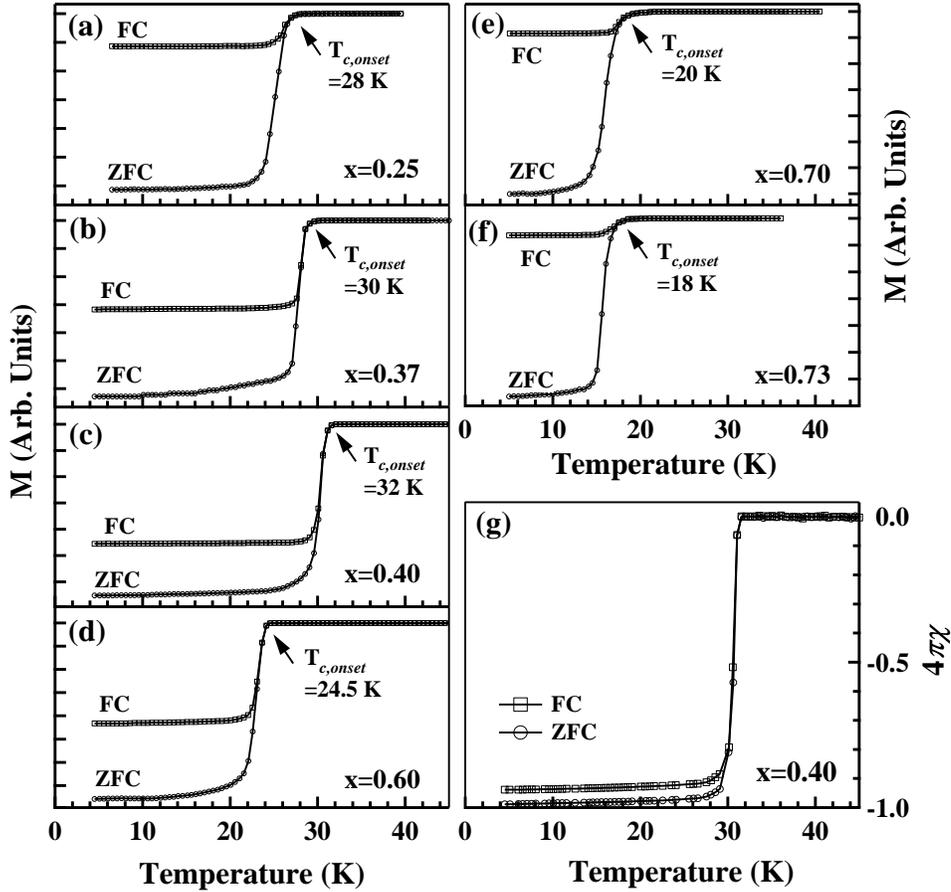}
\end{center}
\caption{Temperature dependence of magnetization of La-Bi2201 single
crystals measured using under a magnetic field of 1 Oe:  (a)
\emph{x} = 0.25, (b) \emph{x} = 0.37, (c) \emph{x} = 0.40, (d)
\emph{x} = 0.60, (e) \emph{x} = 0.70, (f) \emph{x} = 0.73.(g)
Magnetic susceptibility \emph{$\chi$} for another well-annealed
\emph{x} = 0.40 single crystal. }
\end{figure}

The as-grown single crystals were post-annealed in flowing oxygen at
different temperatures (450 $^\circ$C $\sim$ 700$^\circ$C) for 2
$\sim$ 12 days to adjust the carrier concentration and make the
samples uniform. During the annealing process, the crystals were
embedded in the sintered powders with the same composition to avoid
evaporation of some constituents. After annealing, the samples were
quenched in liquid nitrogen. Compared with the as-grown crystals,
such an annealing process has a weak effect on T$_c$ value for most
of the compositions, except for x=0.84 which, when as-grown, is not
superconducting down to 2 K, but can become superconducting with a
T$_c$ up to 10 K after annealing in oxygen for sufficiently long
time.\\

\section{Results and discussion}

As shown in Fig. 1, large La-Bi2201 single crystals with various
compositions, x, have been successfully grown. For the composition
(\emph{x} = 0.73), single crystal as long as $\sim$40 mm has been
obtained. It is found that the crystal size relies closely on the
starting composition. The viscosity of La-Bi2201 increases with
increasing La concentration, x, which makes the molten zone more
stable. This is probably why it is easier to get large single
crystals at high La concentration (Fig. 1).

After post$-$annealing, the single crystals were characterized in
the composition and crystal structure. The superconducting
properties were investigated by both resistivity and magnetic
susceptibility measurements.

\subsection{Compositional analysis}
The chemical composition of the La$-$Bi2201 crystals was determined
by the induction-coupled plasma atomic emission spectroscopy
(ICP-AES). The results are given in Table 1. The actual composition
of the grown single crystals is very close to their nominal
composition for x=0.37$\sim$0.73 while the deviation of the x=0.25
and 0.84 samples is slightly larger. Overall, this indicates that
the sample homogeneity and composition are well controlled during
the entire single crystal growth process.

\begin{figure}[tbp]
\begin{center}
\includegraphics[width=0.8\columnwidth,angle=0]{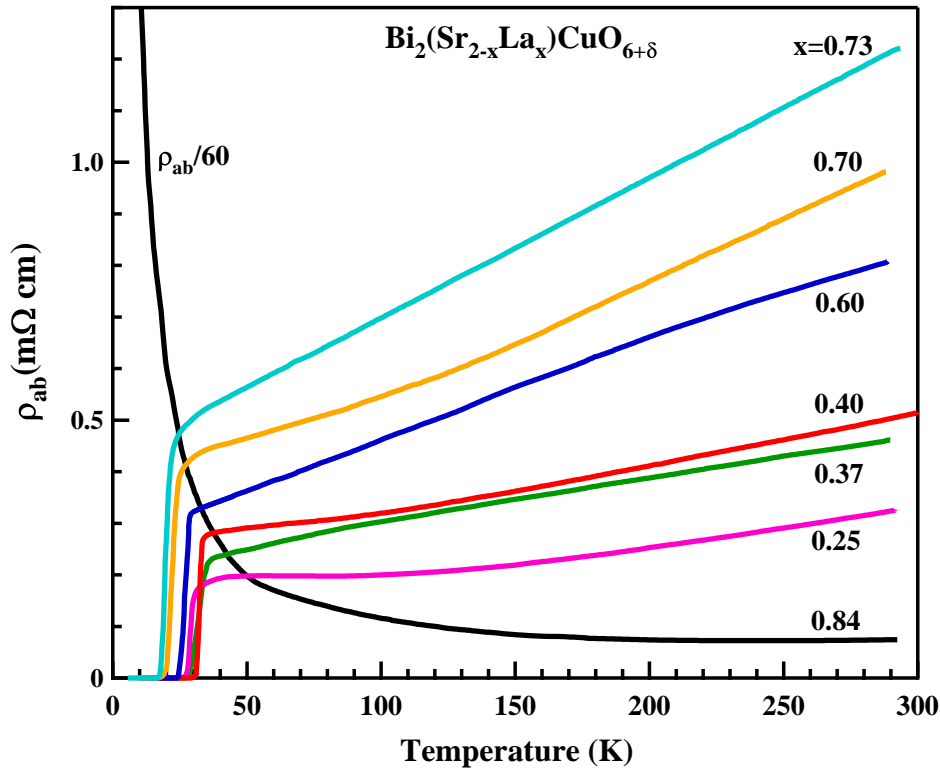}
\end{center}
\caption{Temperature dependence of in$-$plane resistivity
$\rho$$_ab$ of the La-Bi2201 crystals for various nominal La content
(x)
 }
\end{figure}

\subsection{Structure characterization}

The structure and quality of the single crystals are characterized
by X-ray diffraction (XRD) using a rotating anode X-ray
diffractometer with Cu K$_{\alpha}$ radiation ($\lambda$=1.5418
$\AA$). The experiments were carried out under $\theta$-2$\theta$
scan mode with incident ray along the $\emph{c}$-axis of the single
crystal, and the continuous scanning range of 2$\theta$ is from
5$^{\circ}$ to 75$^{\circ}$. Fig. 2 shows XRD patterns for the
La$-$Bi2201 with various compositions \emph{x} = 0.25, 0.37, 0.40,
0.60, 0.70, 0.73 and 0.84. All the observed peaks can be indexed to
the Bi2201 structure, indicating a pure single$-$phase of the
obtained single crystal. The peaks are sharp, as exemplified from
the (008) peaks in the inset of Fig. 2 which has a width of
0.08$\sim$0.09 degree (Full Width at Half Maximum), indicating high
crystallinity and high orientation of the single crystals.

The $\emph{c}$-axis lattice parameters of the single crystals
undergoes a systematic change with the La content x, as shown by the
peak position shift in the inset of Fig. 2.  The \emph{c}$-$axis
lattice constant is calculated from XRD data\cite{C.Dong}, as listed
in Table 1 and shown in Fig. 5(b). It decreases with increasing La
doping which is consistent with the fact that La$^{3+}$ has a
smaller ionic radius than Sr$^{2+}$. The \emph{c}-axis lattice
constant shows a linear relation with the nominal La concentration x
which can be approximated as c(nm)=2.472-0.0795x.  In Fig. 5(b), we
also plot the data from other
groups\cite{Schloegl,Khasanova,B.Liang1} which show a good agreement
with each other.

\begin{figure}[tbp]
\begin{center}
\includegraphics[width=0.6\columnwidth,angle=0]{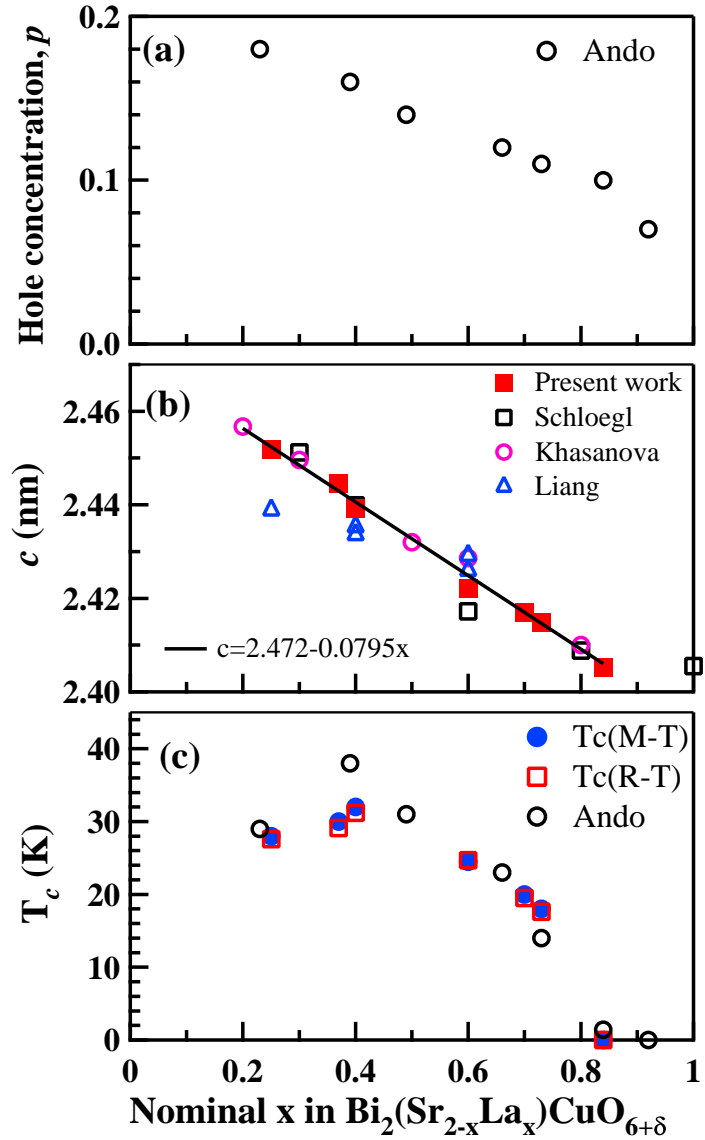}
\end{center}
\caption{(a)Relation of the hole concentration \emph{p} as estimated
from thermopower measurements with the nominal La content by Ono et
al.\cite{S.Ono1}.(b) Dependence of the \emph{c}-axis lattice
constant on the nominal La concentration (\emph{x}). Data from other
groups are also
included\cite{Y.Ando1,Schloegl,Khasanova,B.Liang1,S.Ono2,Y.Ando2}.
The solid line is fitted to our measured data.(c)\emph{T}$_c$ as a
function of nominal La content (\emph{x}). Data by Ono et al. are
also plotted\cite{S.Ono1}.
 }
\end{figure}

\subsection{Superconductivity}
The superconducting transition temperature of the annealed
La$-$Bi2201 crystals is determined by  both magnetization (Fig. 3)
and resistivity (Fig. 4) measurements. Fig.3 shows the temperature
dependence of DC magnetization for La-Bi2201 single crystals with
various La content x, measured using a Quantum Design MPMS XL-1
system with a low magnetic field of 1 Oe. The crystals with x=0.25
$\sim$0.75 show clear superconducting transition while the as-grown
x=0.84 sample is non-superconducting down to 2 K. The
superconducting transition temperature (T$_c$), defined by the onset
of a diamagnetic signal, increases with x first, reaching a maximum
T$_{c,onset}$=32 K near x=0.4, and decreases with further increase
of x. Since the substitution of Sr$^{2+}$ with La$^{3+}$ is expected
to introduce electrons into the samples and it is well-known that
Bi2201 is a hole-doped system, increasing La concentration would
result in a decrease of the hole concentration. Therefore, samples
with x less than 0.4 are expected to be over-doped, x$\sim$0.4 is
optimally-doped while x larger than 0.4 is under-doped.

It is also clear from Fig. 3 that the prepared single crystals show
sharp superconducting transition with a transition width
$\Delta$\emph{T$_c$}=1$\sim$3 K (10$\%$$\sim$90$\%$ standard). In
order to further investigate the quality of the superconducting
samples, we choose a well-annealed x = 0.40 sample for quantitative
investigation. Its magnetic susceptibility has been carefully
calculated by considering  its volume and correcting the
demagnetization factor by considering its shape. As shown in Fig.
3(g), this sample shows a sharp superconducting transition at 31.5 K
with a width of $\sim$1 K.  Its 100$\%$ shielding
(zero-field-cooled, ZFC) signal indicates there exists little
macroscopic inhomogeneity or weak links. Its extremely high Meissner
signal ($\sim$95$\%$, field-cooled£¬FC)  implies that this sample is
nearly free of pinning disorders. Nearly complete shielding and
Meissner effects demonstrate the high quality of this crystal.

Fig. 4 shows the temperature dependence of in$-$plane resistivity
$\rho$$_{ab}$ for all the grown La$-$Bi2201 single crystals with
various compositions. The in$-$plane resistivity was measured using
the standard four-probe method. Gold lead wires were attached to the
contact pads using silver epoxy and then were annealed at
400$^\circ$C. Consistent with DC magnetic measurements (Fig. 3), the
samples with x=0.25$\sim$0.75 show clear superconducting transition
while the x=0.84 sample shows an insulating behavior. The
superconducting transition temperature increases with x first,
reaching a maximum of 31 K (zero resistance) at x=0.40, and getting
smaller with further increase of x. While there may be a relatively
large uncertainty in the absolute value of resistivity, the overall
trend is that the magnitude of $\rho$$_{ab}$ increases with
increasing La content from \emph{x} = 0.25 to 0.84, consistent with
previous measurements\cite{Y.Ando1,S.Ono2,Y.Ando2}.

Fig. 5 summaries the \emph{c}-axis lattice (Fig. 5(b)),
superconducting transition temperature (Fig. 5(c)) of the
La$-$Bi2201 single crystals as a function of La concentration x.
Data from other
groups\cite{Y.Ando1,Schloegl,Khasanova,B.Liang1,S.Ono2,Y.Ando2} are
also included for comparison and completeness. Particularly, the
hole concentration estimated from thermopower measurements by Ono
and Ando\cite{S.Ono1} is plotted in Fig. 5(a). These results provide
useful information in corresponding the carrier concentration,
c-axis lattice constant and the superconducting transition
temperature in La$-$Bi2201 system.

In summary, high-quality and large La$-$Bi2201 single crystals with
wide range of compositions have been successfully prepared by
traveling solvent floating zone method. It will provide an ideal
candidate for investigating the electronic structure\cite{JQMeng}
and physical properties, and superconductivity mechanism of high
temperature superconductors.

\begin{flushleft}
\textbf{Acknowledgments}\\
This work is supported by the NSFC (Project No: 10525417, 10734120,
10874211), the MOST of China (973 project No: 2006CB601002 and
2006CB921302), and CAS (Projects ITSNEM).
\end{flushleft}


\end{document}